\documentclass[aps,floatfix,twocolumn]{revtex4}
\usepackage{latexsym,amsmath,amssymb,amsbsy}
\usepackage{graphicx}
\usepackage{enumerate}

\usepackage{nccfoots}
\usepackage[symbol]{footmisc}

\begin{document}

\title{Charge Topology of the Coherent Dissociation \\ of Relativistic $^{11}$C and $^{12}$N Nuclei}

\author{\textbf{D.~A.~Artemenkov$^{\textbf{1)}}$, V.~Bradnova$^{\textbf{1)}}$, A.~A.~Zaitsev$^{\textbf{1)}}$, P.~I.~Zarubin$^{\textbf{1)~*}}$, I.~G.~Zarubina$^{\textbf{1)}}$, R.~R.~Kattabekov$^{\textbf{1), 2)}}$, N.~K.~Kornegrutsa$^{\textbf{1)}}$, K.~Z.~Mamatkulov$^{\textbf{1), 3)}}$, P.~A.~Rukoyatkin$^{\textbf{1)}}$, V.~V.~Rusakova$^{\textbf{1)}}$, R.~Stanoeva$^{\textbf{4)}}$.}
\Footnotetext{1)}{Joint Institute for Nuclear Research, Joliot-Curie 6, Dubna, Moscow region,141980, Russia.}
\Footnotetext{2)}{Physical-Technical Institute, Uzbek Academy of Sciences, ul. Mavlyanova 2, Tashkent, 700084 Republic of Uzbekistan.}
\Footnotetext{3)}{A.~Kodiriy Jizzakh State Pedagogical Institute, Sharaf Rashidov pr. 4, Jizzakh City, 130100 Republic of Uzbekistan.}
\Footnotetext{4)}{South-West University, Ivan Michailov st. 66, 2700 Blagoevgrad, Bulgaria.}
\Footnotetext{*}{E-mail:\textbf{zarubin@lhe.jinr.ru}}}

\indent  
\noindent \affiliation{Received January 19, 2015}

\begin{abstract} 
\noindent \textbf{Abstract}—The charge topology of coherent-dissociation events is presented for $^{11}$C and $^{12}$N nuclei of energy 1.2~\textit{A}~GeV per nucleon bombarding nuclear track emulsions. This topology is compared with respective data for $^{7}$Be, $^{8,10}$B, $^{9,10}$C and $^{14}$N nuclei. 

\vspace*{16pt}
\noindent \textbf{DOI:}  10.1134$/$S1063778815060022
\end{abstract} 

\maketitle

\section{Introduction}

\indent Light nuclei can be represented as various superpositions of bound states of lighter nuclear cores, extremely light nuclear clusters (alpha particles, tritons, $^3$He nuclei, and deuterons), and nucleons that coexist in dynamical equilibrium. Owing to this variety, the group of nuclei at the beginning of the table of isotopes provides a laboratory for studying the coexistence and evolution of cluster and shell degrees of freedom. The $^{11}$C nucleus, which exhibits a remarkable combination of cluster and shell features of the ground state, exemplify such nuclei. The isotope $^{11}$C is a connecting link between light stable nuclei featuring a pronounced alpha-particle clustering of nucleons and light nuclei at the proton drip line, where clustering that involves the isotope $^3$He is of importance. The interaction of virtual helium isotopes and neutron exchange between them in the $^{11}$C nucleus lead to the formation of cluster structures, including the 2$^{4}$He + $^{3}$He configuration. Among them, configurations characterized by low binding energies, such as $^{7}$Be + $\alpha$ (7.6 MeV), $^{10}$B + \textit{p} (8.7 MeV) and $^{3}$He + $^{8}$Be (9.2 MeV), are expected to be more probable than the 9$^{9}$Be + 2\textit{p} (15.3 MeV) and $^{8}$B + \textit{t} (27.2 MeV) configurations.

\indent Because of a wide variety of virtual cluster configurations that may exist in the structure of the $^{11}$C nucleus, investigation of this nucleus becomes an interesting problem in and of itself.

\indent A balanced coexistence of these virtual cluster modes determines not only the ground-state properties of the $^{11}$C nucleus but also the fact that it is bound, which is of importance for obtaining deeper insight into the abundances of light isotopes. Nuclear astrophysical synthesis of the isotope $^{11}$C may proceed in a mixture of the isotopes ${}^{3}$He and ${}^{4}$He via the formation of the stable isotope $^{7}$Be or the unstable isotope $^{8}$Be, and a partial clustering into a $^{10}$B + \textit{p} pair may follow it. The decay of the $^{11}$C nucleus leads to the formation of the stable isotope $^{11}$B, which one can observe in cosmic rays. This scenario of nucleosynthesis is not commonly recognized-- the isotopes $^{10,11}$B are assumed to be products of the bombardment of carbon-star surfaces with high-energy protons. If observations reveal the dissociation of $^{11}$C nuclei through the ${}^{7}$Be + $\alpha $ and ${}^{3}$He + ${}^{8}$Be channels, this would confirm the existence of cluster modes in this nucleus that are genetically related to its synthesis.

\indent Knowledge of the structure of $^{11}$C is necessary for interpreting data on the next isotope $^{12}$N and, in prospects, on the isotope $^{13}$O, the $^{11}$C nucleus playing the role of a core in both of them. In fast nucleosynthesis processes (hot breakout cycles), these three isotopes are genetically related waiting stations. The formation of the isotope $^{12}$C and heavier nuclei may proceed through them via the addition of protons. It is necessary to know fundamental properties of the relativistic fragmentation of the $^{11}$C nucleus in order to apply intense beams of these nuclei in nuclear medicine.

\section{POTENTIAL OF THE NUCLEAR-EMULSION METHOD}

\indent Within the BECQUEREL project at the nuclotron of the Joint Institute for Nuclear Research (JINR) [1], the cluster structure of light nuclei is being studied in relativistic-fragmentation processes on the basis of the nuclear-emulsion method [2–11]. The development of these investigations and their illustrations are presented in the review article of Zarubin [12]. Among events of the fragmentation of relativistic nuclei, those of their coherent dissociation to narrow jets of fragments are especially important for studying nucleon clustering. They do not feature tracks of either slow fragments of emulsion nuclei or charged mesons. This special feature reflects the fact that the excitation of the relativistic nucleus under investigation is minimal in the case of a tangential collision with a heavy track-emulsion nucleus. Nuclear diffraction interaction [13] not accompanied by angular-momentum transfer is a basic mechanism of excitation of coherent dissociation in nuclear track emulsions. 

\indent The experimental method in question is based on record spatial resolution and sensitivity of nuclear track emulsions whose layers are exposed longitudinally to beams of relativistic nuclei. It has already furnished unique information about cluster aspects of the structure of the whole family of light nuclei, including radioactive ones. For practical reasons, the $^{11}$C nucleus, which is among the key ones, was skipped. A new series of investigations of the BECQUEREL Collaboration was motivated by the need for filling this gap. 

\indent Because of the absence of tracks of strongly ionizing particles, events of coherent dissociation were called white stars. The term white stars reflects successfully a sharp "breakdown" of the ionization density at the interaction vertex upon going over from the primary-nucleus track to a narrow cone of secondary tracks. This special feature generates a fundamental problem for electronic methods because more difficulties should be overcome in detecting events where the degree of dissociation is higher. On the contrary, such events in nuclear-track emulsions are observed and interpreted in the most straightforward way, and their distribution among interaction channels characterized by different compositions of charged fragments is determined exhaustively. This probabilistic distribution is a basic feature that is observed for the virtual cluster structure of the nucleus under consideration. 

\indent The probability distribution of the final configurations of fragments in white stars makes it possible to reveal their contributions to the structure of nuclei under consideration. We assumed that, in the case of dissociation, specific configurations arise at random (random-phase approximation) without sampling and that the dissociation mechanism itself does not lead to the sampling of such states via angular-momentum or isospin exchange. By and large, available results confirm the assumption that cluster features of light nuclei determine the picture of their relativistic dissociation. At the same time, events that involve the dissociation of deeply bound cluster states and which cannot arise at low collision energies are detected. 

\indent We mentioned above that, for the ${}^{11}$C nucleus, one expects the ${}^{7}$Be + $\alpha $, ${}^{10}$B + \textit{p}, and ${}^{3}$He + ${}^{8}$Be dissociation channels, where the binding energies have low values. From the experimental point of view, the last channel is a three-body one and can involve the decays of both the ground state (0${}^{+}$) of the ${}^{8}$Be nucleus and its 2${}^{+}$ excited state. Moreover, channels that, in charge topology, correspond to the dissociation of the ${}^{7}$Be and ${}^{10}$B core nuclei should appear. Thereby, one expects that the role of multi-particle channels in the coherent cluster dissociation of $^{11}$C nuclei should be significant, so that the application of the nuclear-track-emulsion method is reasonable.

\indent In addition, the nuclear-emulsion method should reveal multi-particle channels corresponding in charge topology to the coherent dissociation of the $^{7}$Be [3,11] and $^{10}$B [14] core nuclei in $^{11}$C. An approximate equality of the probabilities for the He + He and He + 2H dominant channels of coherent dissociation is a special feature of the $^{7}$Be nucleus (see Table 1). The respective branching ratio is 1 $\pm$ 0.2 according to data of a group from the Lebedev Physical Institute Moscow [3] and 0.7 ± 0.1 according to vaster data sample obtained at JINR [11]. The 2He + H three-body channels are leading ones(of weight about 75$\%$) among $^{10}$B white stars (see Table 2). Events corresponding to the He + 3H channel saturate 12\%. Lithium and helium fragments appear simultaneously in 10\% of events. Beryllium and hydrogen fragments appear only in 2\% of events; this indicates that, in the structure of the $^{10}$B nucleus, the probability for the $^{9}$Be + \textit{p} configuration is insignificant. On the contrary, the contribution of the Be + Í channel to the coherent dissociation of the $^{8}$B nucleus is dominant, which indicates that the $^{8}$B nucleus features the $^{7}$Be + \textit{p} configuration, which contains a proton halo. The contribution of configurations that involve only helium and hydrogen clusters is estimated at a level of 50\%. 

\begin{table}
\caption{\label{Table:1} Distribution of white stars produced by $^{7}$Be nuclei of energy 1.2 GeV per nucleon among charge channels of nuclear dissociation.}
\begin{center}
\begin{tabular}{|c|c|c|} \hline 
~ Channel ~&~ ${}^{7}$Be [3] ~&~ ${}^{7}$Be [11]~ \\ \hline 
~ 2He ~&~ 41 (44 \%) ~&~ 115 (40 \%)~ \\ \hline 
~ He + 2H ~&~ 42 (45 \%) ~&~ 157 (54 \%)~ \\ \hline 
~ Li + H ~&~ 9 (10 \%) ~&~ 14 (5 \%)~ \\ \hline 
~ 4H ~&~ 2 (2 \%) ~&~ 3 (1 \%)~ \\ \hline 
\end{tabular}
\end{center}
\end{table}

\section{EXPOSURES OF TRACK EMULSIONS}

\indent In December 2013, a set of test samples of nuclear-track emulsions produced at the MICRON workshop of the OJSC Slavich Company was exposed to a secondary beam of relativistic $^{11}$C nuclei at the JINR nuclotron [15]. The samples were prepared by pouring emulsion layers about 200 $\mu$m in thickness onto glass substrates 9$\times$12 cm in size. This nuclear-track emulsion is close in properties to the BR-2 nuclear-track emulsion,which provided sensitivity up to relativistic particles. \par
\indent Nuclei of $^{11}$C were produced in the fragmentation of $^{12}$C nuclei at an energy of 1.2 GeV per nucleon on a polyethylene target 1.5 g cm$^{-2}$ thick. A secondary beam of $^{11}$C nuclei is formed via separation in the magneto-optical channel for beam transportation, the momentum acceptance being about 2\%. At an intensity of the $^{12}$C primary beam on the order of $10^{7}$ nuclei per cycle, the intensity of the $^{11}$C beam is $10^{4}$; this is an optimum value for a controlled exposure of a track-emulsion stack. The beam profile was formed in such a way as to ensure the highest possible degree of uniformity of irradiation along the narrower side.\par
\indent The flux of nuclei that was directed to the irradiated track-emulsion stack was controlled by means of a scintillation monitor. The presence of accompanying nuclei in the composition of the main beam makes it possible to assess the potential of the magneto-optical channel used for the separation of $^{11}$C nuclei [16]. The momentum acceptance of the channel was about 2\%. Figure 1 shows the spectrum of the charge-to-digital converter in transmitting $^{12}$C nuclei. One can observe a contribution of lighter nuclei produced on a generating target at the beginning of the separation channel, the charge-to-mass number ratio being identical to that for the $^{12}$C nucleus. The shoulder on the left of the main peak corresponds to $^{10}$B nuclei, Be nuclei do not manifest themselves ($^{8}$Be is an unbound nucleus), the contribution of lithium nuclei is distinguishable, and helium nuclei manifest themselves quite distinctly.\par
\indent Figure 2 presents the analogous spectrum for the case of tuning the channel to the separation of $^{11}$C nuclei that have the same energy per nucleon as $^{12}$C nuclei. Signals from boron, beryllium ($^{7}$Be), lithium, and helium are only slightly seen and are associated with the fragmentation of $^{11}$C nuclei. The disappearance of helium nuclei is worthy of special note. Upon a decrease in the magnetic rigidity of the channel, $^{4}$He nuclei have already disappeared, while $^{3}$He nuclei have not yet appeared. All these facts indicate that the degree of the separation of $^{11}$C nuclei is quite high, which gives sufficient grounds to disregard the contribution of other isotopes. An irradiation of 40 track-emulsion layers was performed in a beam of this composition. In order to vary the irradiation density, the layers in question were combined into seven stacks irradiated successively.\par
\indent A reduced thickness and glass substrates of the test track-emulsion series turned out to be factors that prevented an analysis that would involve tracing beam and secondary tracks without sampling. Therefore, we scanned track-emulsion layers along transverse bands with the aim of finding tracks of relativistic fragments whose total charge is not less than three and subsequently tracing them up to interaction vertices. Tracks corresponding to singly and doubly charged relativistic fragments were determined visually. The fact that carbon nuclei were dominant in the beam made it possible to establish the charges of heavier fragments in white stars as the values that were needed to reach six charge units.\par 

\section{STATUS OF INVESTIGATIONS OF THE $^{11}$C NUCLEUS} 

\indent In six scanned track-emulsion layers, 144 white stars where the total charge of relativistic fragments is six charge units have been found to date. Their distribution in charge states is given in Table 3. This table also present data on the isotopes $^{10}$C [9] and $^{9}$C [5]. These data indicate that white stars have an individual character for each isotope and that the exposures in our present experiments correspond to the mass numbers of the aforementioned carbon isotopes. In the investigation of the coherent dissociation of relativistic $^{12}$C nuclei in [17], all of the 100 white stars found there arose in the single channel $^{12}$C $\rightarrow ^{3}$He, clearly reflecting a virtual alpha-particle clustering of this nucleus. The discovery of the decay of relativistic $^{8}$Âå nuclei, which made a contribution of magnitude not less than 20\%, was a key observation.

\indent Events featuring only relativistic helium and hydrogen isotopes, especially 2He + 2H, were dominant among $^{11}$C white stars, their weight being as high as 77\%. The branching ratio for the 2He + 2H and He + 4H channels is 6 ± 3, which is at odds with the above idea that only the $^{7}$Be core nucleus undergoes dissociation. 

\begin{center}
\begin{figure}
\includegraphics[width=3.2in]{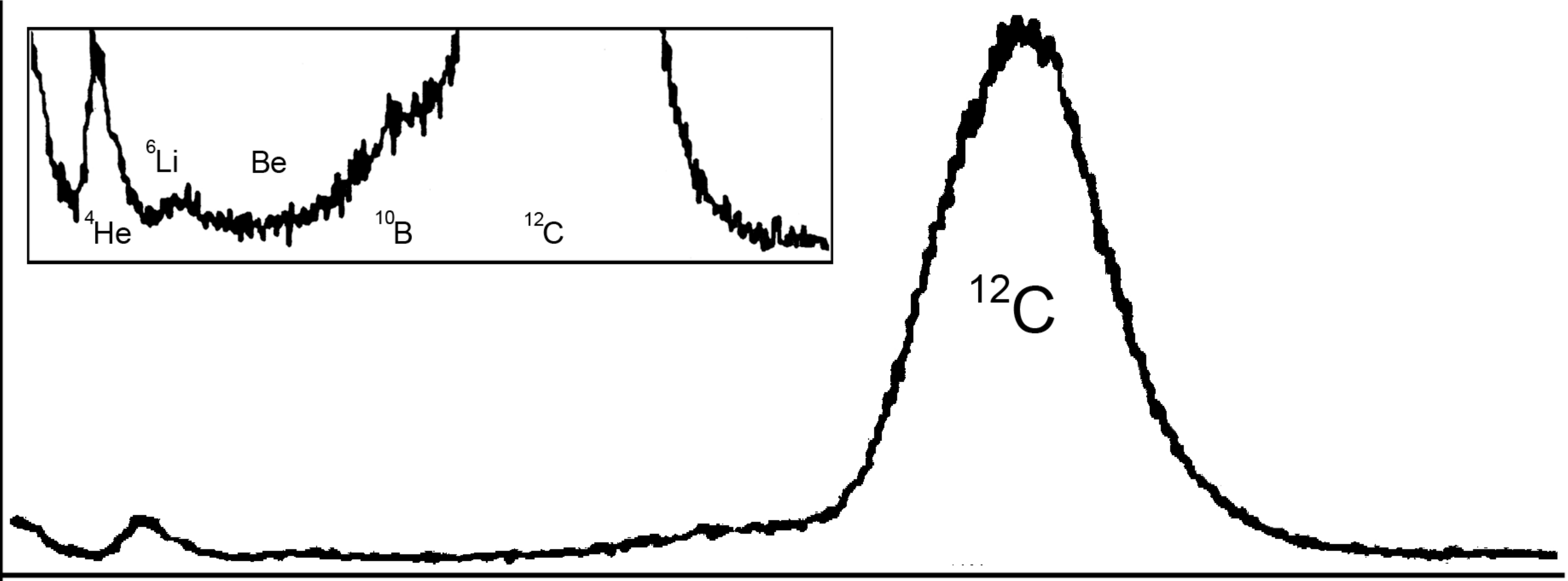}
\caption{Photograph of the spectrum of the charge-to-digital converter of the scintillation monitor for beam nuclei in transmitting $^{12}$C nuclei (arbitrary units). The inset shows the lower left section of the spectrum on an enlarged logarithmic scale.}
\label{fig:1}
\end{figure} 
\end{center} 

\indent In contrast to what was found for the neutron-deficient nuclei studied earlier, we observed here a significant fraction of Li + He + H events, which could correspond to the $^{6}$Li + $^{4}$He + \textit{p} structure. There were no Be + 2H events ( $^{9}$Be + 2\textit{p} cluster structure). At the same time, we observed a significant fraction of Be + He events. In the case of identifying the isotope $^{4}$He in them, one determines unambiguously the isotope $^{7}$Be. Most probably, the 3He channel corresponds to the 2$^{4}$He + $^{3}$He configuration, which may arise both from the decay of the $^{8}$Be and $^{7}$Be core nuclei and from the decay of three-body states. An additional contribution to multi-particle channels may come from the dissociation of the $^{6}$Li cluster as an individual element of the $^{11}$C nucleus in accordance with its virtual $\alpha + d$ structure [18]. Figuratively speaking, we can state that the charge-topology distributions presented above have an individual character for $^{11}$C, which distinguishes it among other isotopes, appearing to be some kind of a signature of this nucleus.

\begin{table}
\caption{\label{Table:2} Distribution of white stars produced by ${}^{10}$B (of energy 1.0 GeV per nucleon) and ${}^{8}$B (of energy 1.2 GeV per nucleon) nuclei among charge channels of the dissociation of nuclei}
\begin{center}
\begin{tabular}{|c|c|c|} \hline 
~ Channel  ~&~ $^{10}$B [14] ~&~ $^{8}$B [4]  ~ \\ \hline 
~ Be + H  ~&~ 1 (2 \%) ~&~ 25 (48 \%)  ~   \\ \hline 
~ 2He + H  ~&~ 30 (73 \%) ~&~ 14 (27 \%)  ~   \\ \hline 
~ He + 3H  ~&~ 5 (12 \%) ~&~ 12 (23 \%)  ~   \\ \hline 
~ Li + He  ~&~ 5 (13 \%) ~&  ~   \\ \hline 
\end{tabular}
\end{center}
\end{table}

\begin{figure}
\includegraphics[width=3.2in]{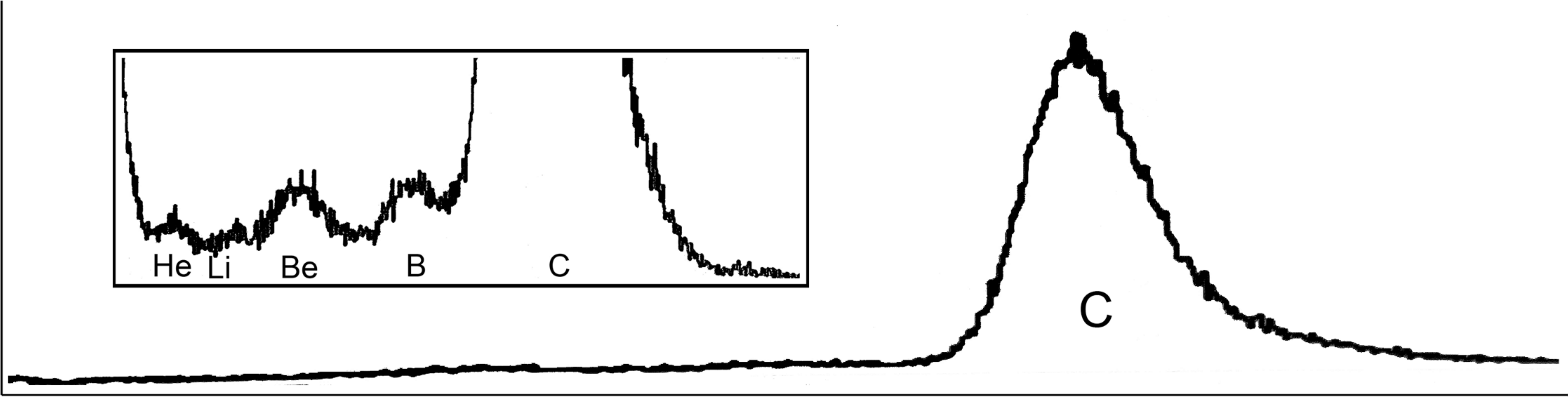}
\caption{As in Fig. 1, but in transmitting $^{11}$C nuclei.}
\label{fig:2}
\end{figure} 

\indent By and large, the structure of $^{11}$C can be thought to be a superposition of cluster states featuring $^{10}$B and $^{7}$Be core nuclei. This prescribes further lines of our investigations. The above aspects furnish a motivation for a new series of studies on the basis of irradiations already performed. They will be aimed at enlarging the statistics of $^{11}$C white stars, as well as performing measurements of multiple scattering in order to identify helium and hydrogen isotopes and angular measurements in order to determine the fraction of 8 Be decays and to explore dissociation dynamics. A selection of values of the total transverse momentum of relativistic fragments within the range characteristic of diffractive dissociation would make it possible to compensate indirectly for the impossibility of directly identifying isotopes heavier than helium. 

\section{INTERPRETATION OF THE CLUSTER STRUCTURE OF THE $^{12}$N NUCLEUS}

\indent The analysis of the experimental data for $^{11}$C nuclei in the preceding section creates preconditions for giving a more justified interpretation of the charge topology of white stars generated by relativistic $^{12}$N nuclei [10]. In Table 4, their statistics are presented along with comparable data on $^{14}$N white stars [19]. The $^{11}$C + \textit{p} (0.6 MeV), $^{8}$B + $^{4}$He (8.0 MeV), and \textit{p} + $^{7}$Be + $^{4}$He channels can make substantial contributions to $^{12}$N white stars. Also, a multi-particle dissociation through the $^{3}$He + $^{9}$B (10 MeV) channel involving an unbound nucleus is possible. An interpretation of the B + 2H channel is complicated by the presence of the $^{10}$B + 2\textit{p} (9.2 MeV) channel. In just the same way as in the case of $^{11}$C, multi-particle channels may emerge owing both to the dissociation of the $^{10}$B core nucleus and to the dissociation of $^{7}$Be. Possibly, a leading character of the 2He + 3H channel (see Table 3) reflects the dissociation of $^{11}$C via a process involving $^{10}$B. A small probability for the dissociation of $^{10}$B to a $^{9}$Be + \textit{p} pair makes it possible to specify Be in Table 4 as $^{7}$Be. Owing to the limit on the mass number, we can specify B from the Be + He channel in Table 4 as $^{8}$B. 

\begin{table}
\caption{\label{Table:3}  Distribution of white stars produced by carbon isotopes (of energy 1.2 GeV per nucleon) among charge channels of the dissociation of nuclei.}

\begin{center}
\begin{tabular}{|c|c|c|c|} \hline 
~  Channel ~ & ~  ${}^{11}$C  ~ & ~  ${}^{10}$C [9] ~ & ~  ${}^{9}$C [5] \\ \hline 
~ B + H ~ & ~  6  (5 \%) ~ & ~  1  (0.4 \%) ~ & ~  15  (14 \%)~   \\ \hline 
~ Be + He ~ & ~  17  (14 \%) ~ & ~  6  (2.6 \%) ~ & ~   \\ \hline 
~ Be + 2H ~ & ~   ~ & ~   ~ & ~  16  (15 \%) \\ \hline 
~ 3He ~ & ~  22  (18 \%) ~ & ~  12  (5.3 \%) ~ & ~  16  (15 \%) \\ \hline 
~ 2He + 2H ~ & ~  60  (48 \%) ~ & ~  186  (82 \%) ~ & ~  24  (23 \%) \\ \hline 
~ He + 4H ~ & ~  14 (11 \%) ~ & ~  12  (5.3 \%) ~ & ~  28  (27 \%) \\ \hline 
~ Li + He + H ~ & ~  4  (3 \%) ~ & ~   ~ & ~   \\ \hline 
~ Li + 3H ~ & ~   ~ & ~  1  (0.4 \%) ~ & ~  2  (2 \%) \\ \hline 
~ 6H ~ & ~  3  (2 \%) ~ & ~  9  (4 \%) ~ & ~  6  (6 \%) \\ \hline 
\end{tabular}
\end{center}
\end{table}

\begin{table}
\caption{\label{Table:4} Distribution of white stars produced by $^{14}$N (of energy 1.2 GeV per nucleon) and $^{14}$N (of energy 2 GeV per nucleon) nuclei among charge channels of the dissociation of nuclei}
\begin{center}
\begin{tabular}{|c|c|c|} \hline 
~  Channel ~ & ~  ${}^{12}$N [10] ~ & ~  ${}^{14}$N [19] \\ \hline 
~  C + H ~ & ~  4  (6 \%) ~ & ~  13 (28 \%) \\ \hline 
~  B + He ~ & ~  3 (4 \%) ~ & ~  4  (9 \%) \\ \hline 
~  B + 2H ~ & ~  11 (15 \%) ~ & ~  3  (7 \%) \\ \hline 
~  Be + He + H ~ & ~  9 (13 \%) ~ & ~  1  (2 \%) \\ \hline 
~  Be + 3H ~ & ~  10  (14 \%) ~ & ~   \\ \hline 
~  Li + He + 2H ~ & ~   ~ & ~  1  (2 \%) \\ \hline 
~  Li  + 4H ~ & ~   ~ & ~  1  (2 \%) \\ \hline 
~  3He + H ~ & ~  2  (3 \%) ~ & ~  17 (37 \%) \\ \hline 
~  2He + 3H ~ & ~  24  (33 \%) ~ & ~  6  (13 \%) \\ \hline 
~  He + 5H ~ & ~  9  (13 \%) ~ & ~   \\ \hline 
\end{tabular}
\end{center}
\end{table}

\indent A limited volume of the statistical sample of $^{12}$N white stars [10] (see Table 4) is due to forming a $^{12}$N beam with the aid of the charge-exchange reaction involving relativistic $^{12}$C nuclei. This way was aimed primarily at simplifying the identification of $^{12}$N white stars on the basis of the total relativistic-fragment charge equal to seven units against a more intense background of events associated with accompanying carbon isotopes. However, a cumbersome determination of charges of beam tracks as those that correspond to seven charge units becomes necessary because of a sizable contribution of events of coherent carbon-isotope dissociation involving meson production within a narrow cone of relativistic fragmentation. This fact drastically reduces the efficiency of the charge-exchange reaction. At the same time, a good separation of carbon isotopes at the JINR nuclotron was confirmed by data on their coherent dissociation. This indicates that, in order to increase sharply statistics of white stars, one can use irradiations of track emulsions with nuclei of the isotopes $^{12,13}$N produced in the fragmentation of relativistic $^{14}$N nuclei. 

\section{CONCLUDING REMARKS}

\indent It has been shown that, in relation to what we have for $^{7}$Be, $^{8,10}$B, $^{9,10}$C, and $^{14}$N nuclei, the track-emulsion method is more efficient in identifying events of the coherent dissociation of relativistic $^{11}$C and $^{12}$N neutron-deficient nuclei. A complete determination of the branching ratios for channels of the breakup of these nuclei makes it possible to reconstruct their virtual cluster structure. The data that we obtained on charge topology is an important first step in this field. 

\indent New possibilities of the track-emulsion method may open upon accelerating $^{16}$O nuclei in order to perform similar investigations in beams of the neutron-deficient isotopes $^{13,14,15}$O. A further advancement toward heavier neutron-deficient isotopes by means of the track-emulsion method remains promising but runs into more serious problems. On this way, the variety of the \textit{p}–-$^{3}$He–-$\alpha$ ensembles under study may become ever wider. 

\indent Starting from the $^{11}$C and $^{12}$N nuclei, there arise problems associated with the limitations of the approach based on the coherent dissociation of relativistic nuclei in nuclear-track emulsions—specifically, we mean here the impossibility of a direct identification in mass number for relativistic fragments heavier than helium. The fraction of events involving such fragments increases sharply as the mass number of the nuclei under study grows. At energies of incident nuclei (Å) around a few GeV units per nucleon, one can identify them in electronic experiments involving a magnetic analysis. In the future, such an identification may become possible for \textit{Å} in the region of several tens of GeV units per nucleon in experiments with hadron calorimeters. Investigations on the basis of the track-emulsion method remain valuable since they may provide guidelines for electronic experiments aimed at studying the coherent dissociation of relativistic neutron-deficient nuclei. 

\begin{center}
ACKNOWLEDGMENTS
\end{center}

\indent We are grateful to A.I.~Malakhov (JINR), N.G.~Polukhina, and S.P.~Kharlamov (Lebedev Physical Institute), and N.S. Zelenskaya (Skobeltsyn Institute of Nuclear Physics, Moscow State University) for their support of this study and critical comments on its results.\par 
\indent This work was supported by the Russian Foundation for Basic Research (project nos. 12-02-00067 and 15-02-01073) and by grants from the plenipotentiaries of Bulgaria, Egypt, Romania, and the Czech Republic at JINR.

\begin{center}
\textbf{REFERENCES}
\end{center}

\begin{enumerate}
\item  The BECQUEREL Project, http://becquerel.jinr.ru/.
\item  D.~A.~Artemenkov \textit{et al.}, Phys. Atom. Nucl. 70, 1222 (2007); nucl-ex/ 0605018.
\item  N.~G.~Peresadko \textit{et al.}, Phys. Atom. Nucl. 70, 1266 (2007); nucl-ex/0605014.
\item  R.~Stanoeva\textit{ et al.}, Phys. Atom. Nucl. 72, 690 (2009); arXiv: 0906.4220. 
\item  D.~O.~Krivenkov\textit{ et al.}, Phys. Atom. Nucl. 73, 2103 (2010); arXiv: 1104.2439. 
\item  R.~R.~Kattabekov\textit{ et al.}, Phys. Atom. Nucl.\textit{ }73, 2110 (2010); arXiv: 1104.5320.
\item  D.~A.~Artemenkov\textit{ et al.}, Few Body Syst. 50, 259 (2011); arXiv: 1105.2374.
\item  D.~A.~Artemenkov\textit{ et al}.,\textit{ }Int. J. Mod. Phys. E 20, 993 (2011); arXiv: 1106.1749.
\item  K.~Z.~Mamatkulov\textit{ et al.}, Phys. At. Nucl. 76, 1224(2013); arXiv: 1309.4241.
\item  R.~R.~Kattabekov\textit{ et al.}, Phys. At. Nucl. 76, 1219(2013); arXiv: 1310.2080.
\item  N.~K.~Kornegrutsa \textit{et al.} Few Body Syst. 55, 1021 (2014); arXiv: 1410.5162.
\item  P.~I.~Zarubin, Lect. Notes in Phys., 875, 51(2013) Springer Int. Publ.; arXiv: 1309.4881.
\item  N.~G.~Peresadko \textit{et al.}, JETP Lett. 88, 75 (2008); arXiv: 1110.2881.
\item  M.~I.~Adamovich, Phys. At. Nucl. 67, 514 (2004); nucl-ex/0301003.
\item  The Slavich Company,  www.slavich.ru, www.newslavich.com.
\item  P.~A.~Rukoyatkin \textit{et al.}, EPJ ST 162, 267(2008); nucl-ex/1210.1540.
\item  V.~V.~Belaga \textit{et al.}, Phys. At. Nucl. \textbf{58}, 1905 (1995)]; nucl-ex/1109.0817.
\item  M.~I.~Adamovich \textit{et al.}, Phys. At. Nucl. \textbf{62}, 1378 (1999); nucl-ex/1109.6422.
\item  T.~V.~Shchedrina \textit{et al.}, Phys. Atom. Nucl. \textbf{70}, 1230 (2007); nucl-ex/0605022.
\end{enumerate}

\end{document}